\documentclass[12pt,preprint]{aastex}
\usepackage{natbib}
\usepackage{graphicx}
\usepackage{amsmath}
\usepackage{epsfig}
\begin{document}
\title{The Foggy Disks Surrounding Herbig Ae Stars: a Theoretical Study of
  the H$_2$O Line Spectra}
\author{J.Cernicharo$^{1,2}$, C. Ceccarelli$^2$, F. M\'enard$^2$,
  C. Pinte$^2$, A. Fuente$^3$}
   \altaffiltext{1}{Laboratory of Molecular Astrophysics.
    CAB (CSIC-INTA). Crta Ajalvir km4. 28840 Madrid. Spain.
     jcernicharo@inta.es}
   \altaffiltext{2}{Laboratoire d'Astrophysique de Grenoble, UMR
5571-CNRS, Universit\'e Joseph Fourier, Grenoble,France}
   \altaffiltext{3}{Observatorio Astron�mico Nacional (OAN), Apdo. 112, 28800 Alcal\'a de Henares, Madrid, Spain}
    \date{Received June 16th , 2009; accepted August 17th , 2009}
\begin{abstract}
  Water is a key species in many astrophysical environments, but it is
  particularly important in proto-planetary disks. So far,
  observations of water in these objects have been scarce, but the
  situation should soon change thanks to the Herschel satellite. We
  report here a theoretical study of the water line spectrum of a
  proto-planetary disk surrounding Ae stars. We show that several
  lines will be observable with the HIFI instrument onboard the
  Herschel Space Observatory. We predict that some maser lines
  could also be observable with ground telescopes and we discuss how
  the predictions depend not only on the adopted physical and chemical
  model but also 
  on the set of collisional coefficients used and 
  on the H$_2$ ortho to para ratio through its effect on
  collisional excitation. This makes the water lines observations a
  powerful, but dangerous -if misused- diagnostic tool.
 \end{abstract}


\section{Introduction}
Water is a very special molecule in the interstellar medium for its
crucial role in cooling the molecular gas. However, because of its
abundance in the Earth's atmosphere, observations of water have to be
made from space. Notable exceptions are the observations of its
isotopes (H$_2^{18}$O, HDO and D$_2$O), whose rotational transitions
can be observed with ground-based telescopes (e.g., \citet{Butner07}
for the detection of D$_2$O, or some maser lines
\citet{Cerni90,Cerni94,Cerni96,Cerni06}).

Similarly, water is of particular importance in protoplanetary disks,
the sites where planets are formed.  More difficult to detect than in
other astrophysical environments, HDO has so far been detected in only
one protoplanetary disk \citep{Cecca05}, and only very
recently the Spitzer Space Telescope detected warm H$_2$O from
the innermost regions of protoplanetary disks \citep{Carr08,
Salyk08}.  These observations rose the question of the
origin of the observed warm water as H$_2$O was predicted to be absent
in the innermost disk.  On the contrary, water is expected
to be present in the outer regions of the disk, where the bulk of the
gas resides. But water has never been detected in those colder
regions. The situation will hopefully soon change thanks to the
Herschel Space Observatory, and, specifically, the HIFI
instrument which will allow to observe the water lines with
unprecedented sensitivity, spectral and spatial resolution. In this
Letter, we report the predictions of the H$_2$O spectrum toward a
typical disk surrounding a Herbig Ae star, similar to the one of
HD~97048. 

We first describe the adopted physical and chemical structures, based
on previously published models, and the method to compute the emerging
line spectrum (\S \ref{sec:model-description}). We then show the
results of our computations (\S \ref{sec:results}) and discuss how
they depend on the distribution of the grain sizes across the disk, a
key parameter in the evolution of the protoplanetary disks and the
formation of planets. We also show the impact of the adopted set of
collisional rates on the predictions and the importance of having a
full modeling of the spectrum to derive reliable physical and chemical
parameters from the water lines.

\section{Model description}\label{sec:model-description}

The disk structure used in the present calculations is inspired by the
disk model derived for HD~97048 by \citet{Lagage06} and \citet{Doucet07}.  
The dust mass in the disk is 10$^{-4}$M$_\odot$,
distributed following power-laws for the scale height and surface
density, and located between R$_{in}=$0.9~AU and R$_{out}=$ 400~AU.
The flaring exponent is $\beta =$ 1.26 and the exponent for the
surface density is $p = -1.5$.  The gas scale height is 51~AU at the
reference radius of 135~AU. The continuum radiative transfer and dust
temperature calculations are performed with MCFOST, a Monte Carlo
based code \citep{Pinte06, Pinte09}.  The central star has
T$_{eff}$=10000K and a radius of 2 R$_\odot$. 
The dust opacity is calculated with the Mie theory
 and the dust composition is the mixture given by $model A$ of 
\citet{Mathis89}. 
The grains size distribution follows a power-law ranging from 
0.03$\mu$m to 1mm in radius and the usual ISM slope of -3.5. 
Two models are calculated, one where dust is fully
mixed throughout the disk, and one where vertical settling has  occurred
and large particles have been removed from the disk surface and
progressively settled to the disk midplane as a function of their
size, the larger the grain the more complete the settling.  In the
following we refer to the latter as stratified disk model.
Details about the parameterization of the dust settling are given in
\citet{Pinte08}. The exponent $\chi_i$ describing the settling has
been fixed to 0.25, a value that fits the observations of GG Tau
\citep{Pinte07}.  The continuum calculations made with MCFOST
provided the density and temperature profiles, as well as the UV
radiation field fully propagated including multiple scattering, to
further compute the H$_2$O abundance and line spectrum with other
tools described below.

\begin{figure}
\includegraphics[scale=0.90]{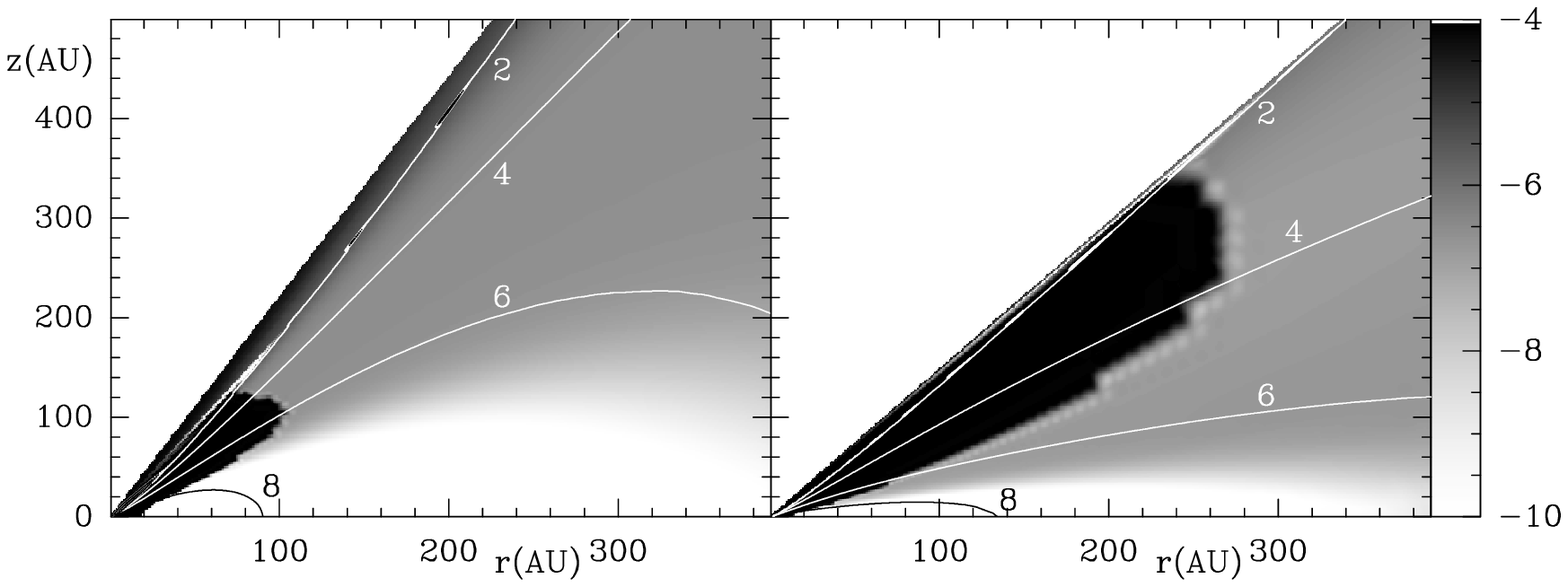}
\caption{Water abundance, relative to H$_2$,
  distribution in the disk surrounding HD~97048,
  in the case of no stratification (left panel) and stratification
  (right panel) of the dust grains (see text). The right bar shows the
  value of the grey contours. The lines mark the H$_2$
  logarithmic density, increased by steps of 100 from 10$^2$ 
  to 10$^{8}$ cm$^{-3}$ (upper \& lower contours respectively). 
\label{fig:chemi} }
\end{figure}

The H$_2$O abundance profile has been computed following the model
described in \citet{Dominik05}. The model has been verified
against the more extended chemical model by \citet{Willacy07} to give
similar results. Briefly, at the conditions prevailing in the
protoplanetary disks described above, the major reservoir of water is
that frozen on the dust grain mantles. The latter inject water into
the gas phase because of two major mechanisms: i) the sublimation of
the H$_2$O-rich ices when the dust temperature exceeds about 100 K,
and ii) the photo-desorption of the H$_2$O-rich ices due to the FUV
photons from the Interstellar Field and the star itself 
\citep{Dominik05, Willacy07}. 
These two mechanisms completely dominate the
abundance profile of the water across the disk.  The results for the
disk described above are shown in Fig. \ref{fig:chemi} which shows the
case of a standard grain size distribution and the case where a
stratification is present (see above). In both cases, three regions
can be identified: 1) at radii less than about 20 AU, water is
abundant at any height of the disk because of the warm dust
temperature ( $\geq100$ K) and water ice sublimation; 2) between 20~AU
and about 100 AU (or 250 AU in the case of dust stratification), water
is mostly frozen onto the grain mantles on the equatorial plane (where
the density is larger than $\sim 10^6-10^{7}$ cm$^{-3}$ and the dust
temperature lower than $\sim100$ K) but it is very abundant ($\sim
1\times10^{-4}$ with respect to H$_2$) in the regions just above the
plane because of ice sublimation; 3) at radii larger than about 100 AU
(or 250 AU in the case of dust stratification), water is frozen in the
equatorial plane and abundant ($\sim 3\times10^{-7}$) above it because
of the photo-desorption of the ices by the FUV photons.  The major
difference in the water abundance distribution between the non
stratified and stratified case is, therefore, a much larger region
where ices sublimate and water is abundant ($\simeq$10$^{-4}$).  We
will show that this has important consequences on the emerging H$_2$O
line spectrum. Another important difference between the two cases is
the continuum emission from the dust grains which also largely affects
the emerging water line profiles.

%


Given the intrinsic geometry of the problem, a full modeling of the
water line spectrum would require a 2-D treatment. However, in the
specific case of water, the lines are greatly optically thick and,
because of the disk geometry, the line optical depths in the
horizontal direction are by far larger than in the vertical one. We
therefore developed a pseudo 2-D code where we computed the radiative
transfer along the vertical direction at 50 and 70 different disk
radii for the non stratified and the stratified case
respectively. Other authors have discussed and verified the validity
of this approach \citep{Pavl07, Asen07}.  
The radial distances for each set of models have been
selected in order to sample the variations of density and water
abundance given by the physical structure of the disk. We have assumed
an intrinsic turbulent velocity of 0.5 kms$^{-1}$ (see, e.g., the
case for MWC758 and CQ Tau published by \citet{Chapillon08}).
The sampling in height has been adjusted to trace narrowly the upper
layers of the disk where the line opacity of the thicker lines is
$\le$1 and photons do escape. The equatorial zone is treated as a
continuum source (water abundance in these regions is $<$10$^{-10}$)
with a temperature identical to that of the dust grains and a spectral
dependence on the dust opacity of 0.65.  The code used for the 1-D
radiative transfer models has been described in \citet{Gonza93, Gonza97}.  
We used the collisional rates with the H$_2$
by \citet{Faure07}, assuming that all H$_2$ is in the ortho
form. We will discuss the impact of this assumption on the computed
emerging line spectrum by comparing the predicted emerging spectrum
obtained by considering different sets of collisional
coefficients. Finally, in all models the orto-to-para H$_2$O ratio was
assumed equal to 3:1.

\section{Results and Discussion}\label{sec:results}

\begin{figure}[bt]
\includegraphics[scale=1.4]{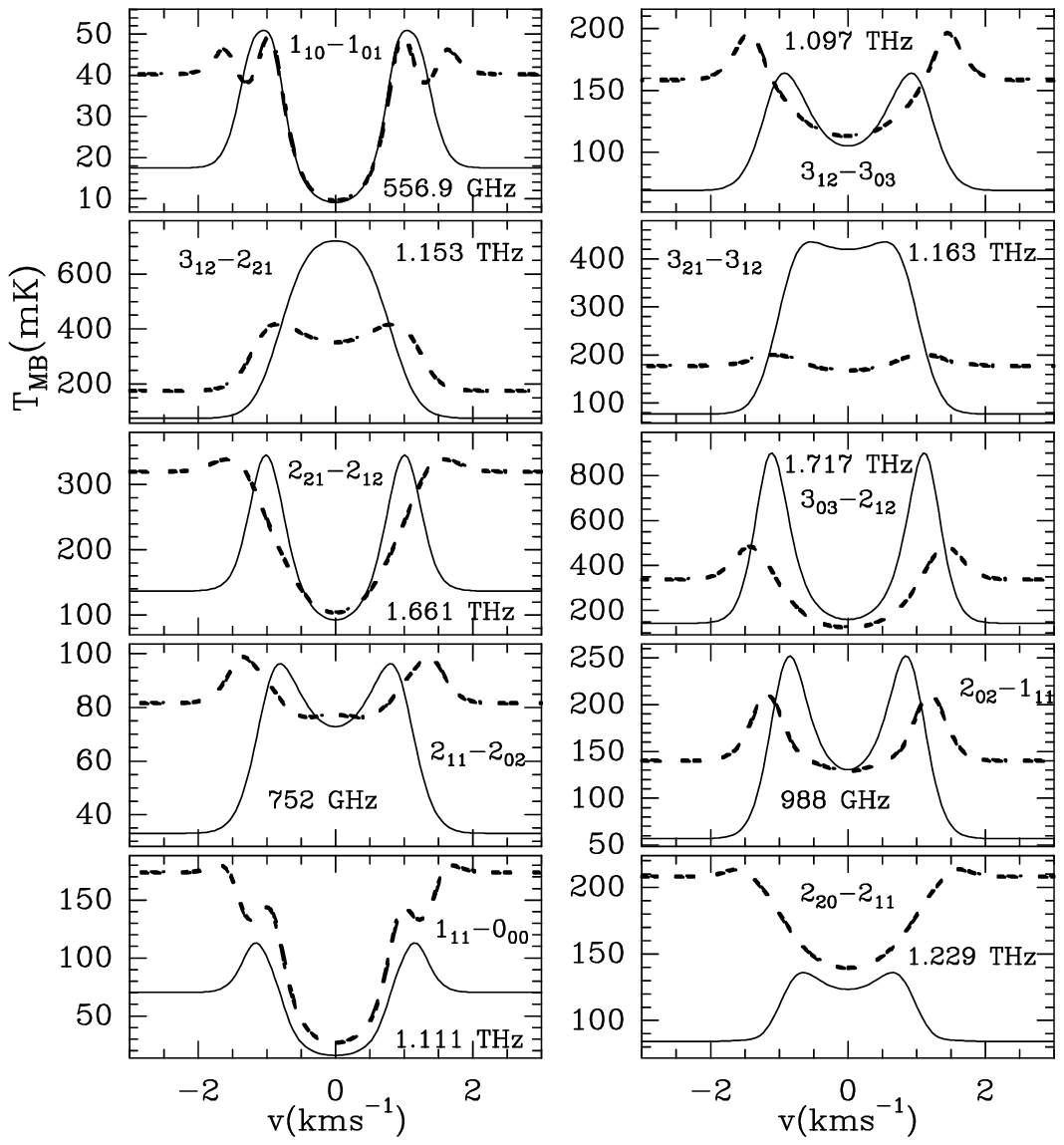}
\caption{The profiles of the eight brightest H$_2$O lines observable
  with the Herschel HIFI spectrometer. The frequency of the line is
  reported in each panel. The ordinates report the T$_{\rm {mb}}$ in
  mK, and the abscissas refer to the velocity in kms$^{-1}$. The
  emerging profiles result from the convolution of the Herschel HIFI beam
  (40$"$ at 557 GHz and 13$"$ at 1.7 THz respectively) with the computed
  brightness temperatures of H$_2$O. Solid lines
  refer to the stratified case, while dashed lines
  refer to the case with no stratification (see text). \label{fig:h2o-lines}}
\end{figure}

\begin{figure}
\includegraphics[scale=1.00]{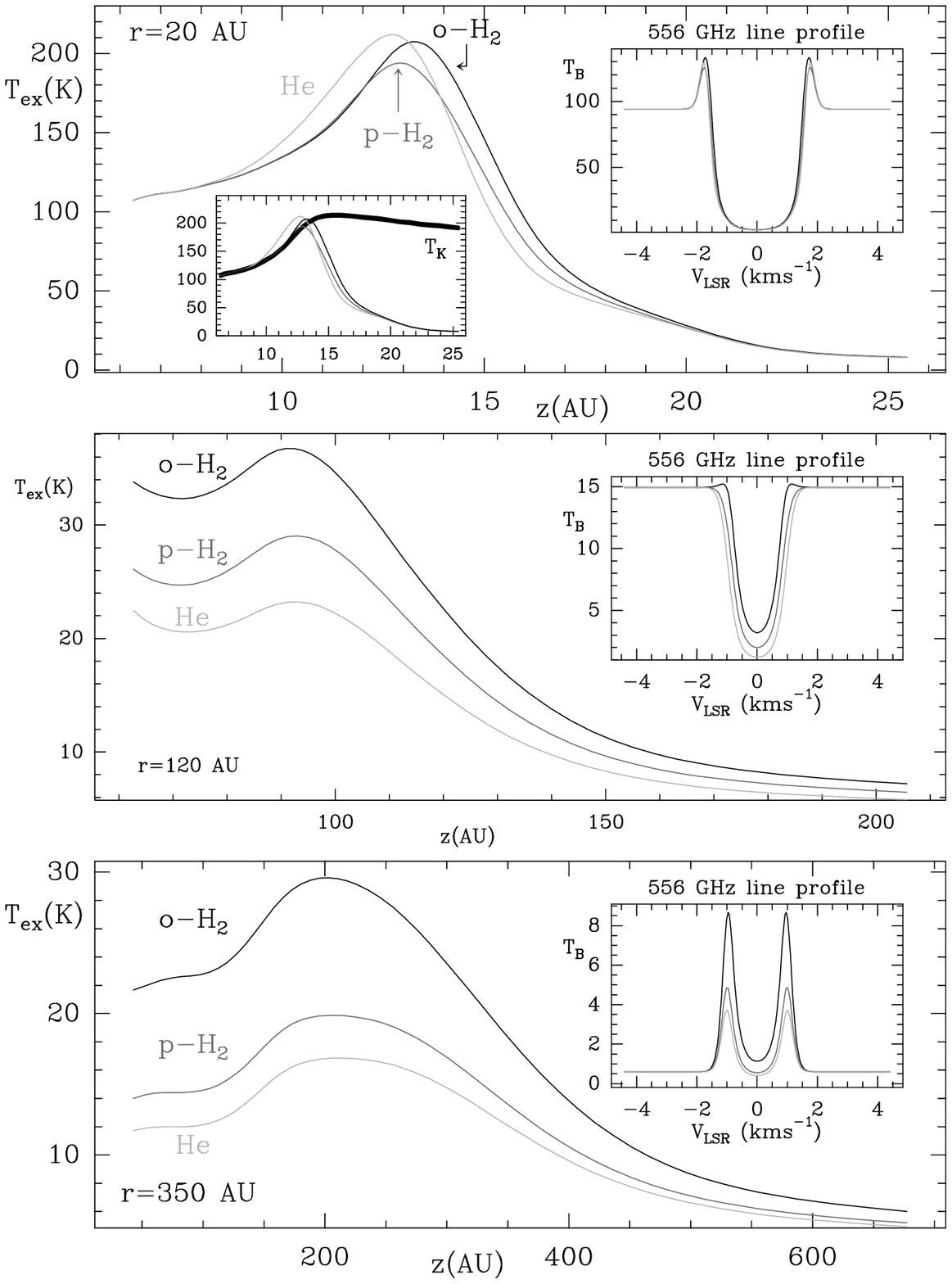}
\caption{Excitation temperature of the fundamental line of water for
radii equal to 20, 120 and 350 AU in function of the height above the 
plane and for
three different sets of collisional rates (see text). The inset in each
panel shows the emerging 1$_{10}$-1$_{01}$ intensity for each set of collisional
rates. The second inset in the top panel shows the excitation temperatures and 
the temperature of the gas. For larger radii the lines are almost subthermally
populated at all heights.
\label{fig:Tex}}
\end{figure}

Figure \ref{fig:h2o-lines} shows the profile of a selected sample of
lines, namely the brightest lines observable with the HIFI instrument
onboard the Herschel observatory. In order to identify the regions
contributing to the emerging profiles we present the results for a
face-on disk convolved with the beam of Herschel's telescope.  Hence,
the line profiles are totally dominated by the opacity of the lines
and the variation of excitation temperature with radius and height
above the disk.  Several kind of profiles are predicted, depending on
the line and whether dust grains are stratified or not. In the case of
dust stratification, several lines are predicted to be in emission
with a central dip caused by self-absorption. Other lines show a
combination of emission and absorption profiles. For example, the para
line at 1.111 THz is in absorption with two ``emission peaks'' at
velocities at about 1 km/s arising from the outer (r$\geq$200 AU)
disk, where the water abundance is lower. Another important example is
represented by the 557 GHz line, which shows two emission peaks again
at about 1 km/s produced by the regions with r$>$180 AU, and a deep
absorption in the central velocities of the continuum produced at all
radii but particularly strong at r$<$ 120 AU (see Fig. \ref{fig:Tex}).
The situation dramatically changes for the case with no dust
stratification. Almost all lines go entirely in absorption against the
dust continuum, because of the brighter continuum. The only exception
is the 1153 GHz line, thanks to its relatively low spontaneous
emission coefficient (10 times lower than, for example, the line at
1163 GHz).  In general, the line profiles are dominated by the line
optical depth and the variation of excitation temperature with height
and radius.  Besides, due to the relatively large telescope beam and
to the large line opacities the water lines of Fig.
\ref{fig:h2o-lines} largely probe the upper layers of the outer disk
where ices are photo-desorbed. The wiggles presented by some
  water lines in the no-stratified model are due to the strong
  decrease of water abundance for r$>$120 AU compared to the
  stratified case.

\begin{figure}
\includegraphics[scale=1.00]{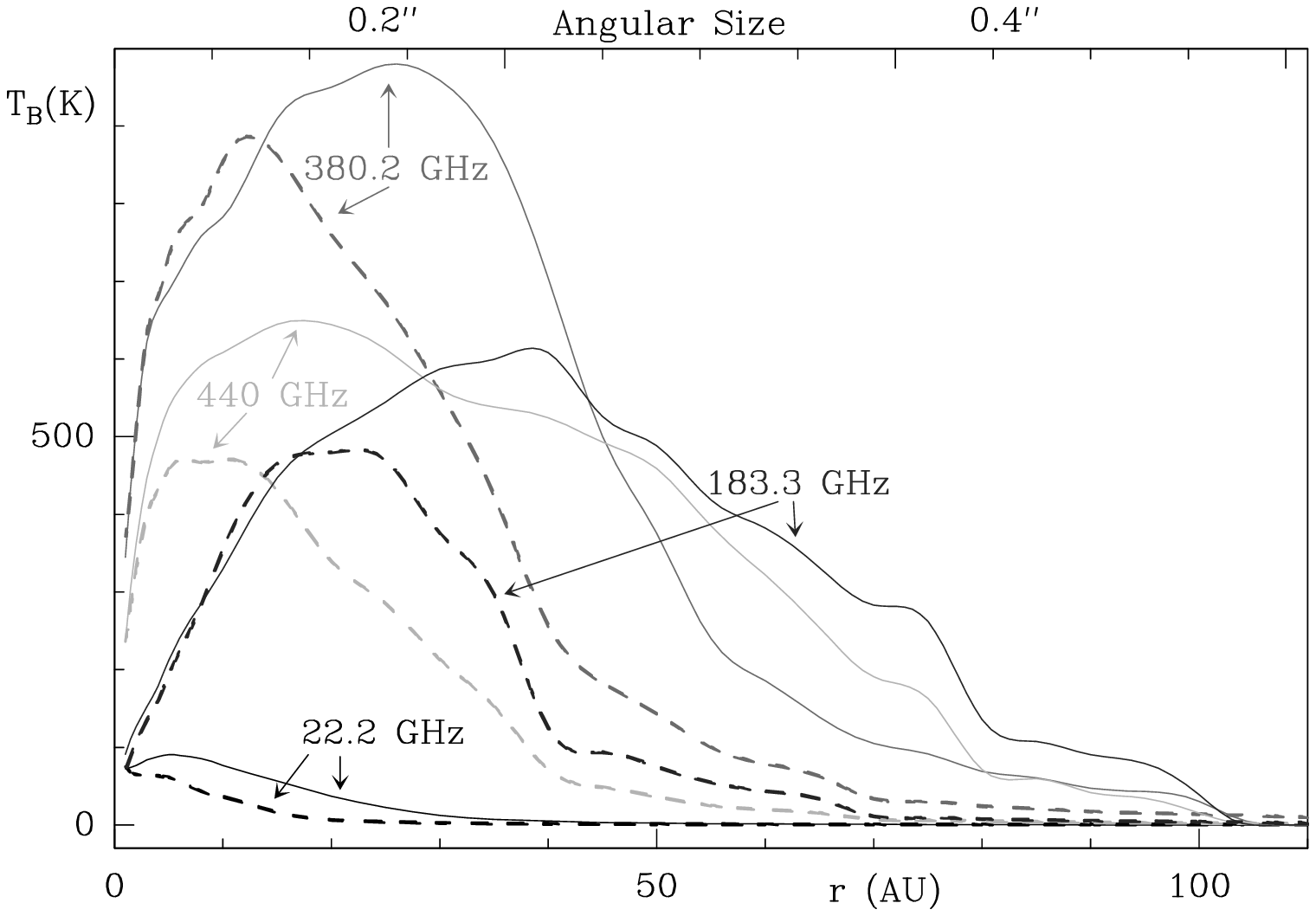}
\caption{Brightness temperature T$_{\rm B}$ (in K) of the four
  brightest maser lines at 22.2, 183.3, 380.2 and 440 GHz respectively
  (as marked on the curves). Dashed lines refer to the case with no
  dust grain stratification, solid lines to the case with
  stratification. The lower x-axis reports the linear distance from
  the star in AU, while the upper x-axis reports the angular distance
  in arcseconds. \label{fig:maser}}
\end{figure}

Several H$_2$O lines show masering effects in the inner disk
equatorial plane because of the high densities and temperatures
prevailing there. Figure \ref{fig:maser} shows the brightness
temperature T$_{\rm B}$ of the four brightest maser lines as a
function of disk radius. All of them probe the inner ($\leq 50$ AU)
and denser regions of the disk, but each of them peaks at a different
radius, probing, therefore, a different part of the disk.  In
principle, observations of these maser lines would allow to constrain
the physical and chemical structure of the inner disk, even if the
spatial resolution is not large enough to resolve the emission.  Of
all the shown maser lines, the 183.3 GHz seems to be the most
promising. For example, we predict a signal of 3 and 1 K at the
IRAM-30m telescope for the non stratified and the stratified case
respectively. Such signals can easily be detected under good weather
conditions \citep{Cerni90}. This is particularly true for the ALMA
interferometer.

We have also computed the emerging profiles for the isotope
H$_2^{18}$O of water assumed to be 500 times less abundant than
H$_2$O.  In spite of the large abundance ratio the intensities
predicted for the fundamental line of the rare isotope are not very
different from those of the main isotope.  Of course, the H$_2^{18}$O
fundamental line penetrates deeper in the disk probing a different
region than the main isotope.
Finally, in order to help understanding the role of collisional
pumping on the emerging water line spectrum, Fig. \ref{fig:Tex} shows
the excitation temperature, T$_{\rm ex}$, of the fundamental line of
water 1$_{10}$-1$_{01}$ as function of the height above the disk at
radii 20, 120 and 350 AU.  Three different sets of collisional
coefficients are used: i) collisions with He, scaled by the H$_2$
mass, as computed by \citet{Green93}; ii) collisions with the
ortho-H$_2$, as computed by \citet{Faure07}; iii) collisions with
para-H$_2$, computed by the same authors.

We notice three effects. First, T$_{\rm ex}$ is always larger for
computations obtained with the ortho-H$_2$ collision coefficients set
and smaller with the He set. This directly reflects the value of the
collisional coefficients, larger for the former set of
coefficients. Second, at the three radii the difference in the T$_{\rm
  ex}$ obtained with the three sets of collisional coefficients is
larger going towards the equatorial plane, and can be as high as a
factor two. This reflects the different population mechanisms at work
in the different regions. At larger heights, the difference diminishes
because the lines are more and more radiatively populated, an effect
nicely seen at radii 20 AU where the three sets of collisional
coefficients produce the same T$_{\rm ex}$. Third, the lines,
integrated along the disk height, at r=120 and 350 AU, show
slightly different profiles and very different intensities,
reflecting the different excitation conditions and line optical
depths. At the smaller radius, 120 AU, the line is in absorption
against the continuum: the absorption is larger in the case of the He
collisional coefficients set and smaller in the case of ortho-H$_2$
set. At the larger radius, 350 AU, the line goes in emission with a
self-absorption dip in the central velocities: the line is weaker in
the case of the He collisional coefficients set and brighter in the
case of ortho-H$_2$ set.

The emerging line profile is obtained by integrating these profiles
over the radii and will, therefore, depend on the balance of the
absorption/emission in the different parts of the disks. In the
specific case of the 556 GHz line reported in Fig.
\ref{fig:h2o-lines}, the line can be substantially in absorption or in
emission, depending on the choice of the collisional coefficients. For
small radii (r$<$120 AU), the line profiles do not depend on the set
of collisional rates while differences of up to a factor of 2 can be
found at larger radii.

We have performed the full calculations for the case in which H$_2$ is
in the para form, rather than ortho as assumed in the calculations of
Fig. \ref{fig:h2o-lines}, using again the collisional coefficient
calculated by \citet{Faure07}. The emerging line profiles dramatically
change: lines in emission become weaker and the absorption becomes
stronger. We note that we obtain similar results (with differences
below 5\%) when we used the more recent collisional coefficients by
\citet{Duber09}.  In summary, the emerging profile of the water
spectrum strongly depends on the o-H$_2$/p-H$_2$ abundance ratio,
which likely depends on the chemical history prior to the formation of
the protoplanetary disk and its subsequent chemical evolution. The
interpretation of the water line profiles will remain, therefore,
challenging in spite of the accurate collisional rates available in
the litterature. 
In summary, the major results of the present work are the following:
{\bf a)} First, and foremost, several H$_2$O lines are predicted to be
detectable with the newly launched space-born telescope Herschel. Some
maser lines are also observable with ground telescopes.
{\bf b)} Even though we did not study the case of a different chemistry (for
example the absence of water ices photodesorption, as in several
previous published models; e.g. \citet{Aikawa06}, or \citet{Woitke09}, 
the comparison between the stratified versus no stratified
case shows that the assumed water abundance profile is crucial in the
emerging line spectrum.  The comparison of our predictions with
  those recently published by Woitke et al. (2009), who used a
  different chemical structure, strengthens this statement. The
  differences between our results and theirs are mainly due to the
  different water abundance profile, in particular in the upper
  layers, to the dust settling we have introduced in our models, and
  to the inclination assumed for their disk.  Besides, the dust
continuum emission is also crucial.  In fact, if not for other
reasons, line fluxes and profiles depend on the position of the water
molecules with respect to the dust absorbing continuum. Therefore,
predictions based on simplistic models based on
constant abundance across the disk cannot be trusted. 
{\bf c)} 
The transition where a line goes from emission to absorption not
only depends on the dust and gas temperature and density profiles, but
also on the collisional coefficients. This dependence is indeed
critical. In practice, the emerging line spectrum strongly depends on
the H$_2$ para-to-orto ratio, a poorly known quantity.
To conclude, observations of water lines will be a very powerful
diagnostic tool to understand the structure of proto-planetary disks.
The analysis of water line profiles will be a challenge because the
levels are sub-thermally populated, the dust photons play a crucial
role in the pumping, and the huge line opacities favour the vertical
diffusion of photons. 
On the other hand, the information
which can be extracted by H$_2$O observations warrants the effort:
amongst others, the amount of water present in the first phases of a
planetary system birth.
\vskip 0.25cm
  This work has been supported by Spanish MICINN 
  through grant AYA2006-14876, by DGU of the CM 
  under IV-PRICIT project S-0505/ESP-0237
  (ASTROCAM). We also thank the French ANR
  (contracts ANR-08-BLAN-0225, ANR-07-BLAN-0221) and
  PNPS of CNRS/INSU for support.  
  We thank J.R. Goicoechea for useful comments.


\begin{thebibliography}{58}
\bibitem[Aikawa \& Nomura, 2006] {Aikawa06} Aikawa Y., Nomura H., 2006, \apj, 642, 115
\bibitem[Asensio-Ramos et al., 2007]{Asen07} Asensio-Ramos A., Ceccarelli C., Elitzur M., 2007, \aap, 471, 187  
\bibitem[Butner et al., 2007]{Butner07} Butner H., Charnley S., Ceccarelli C., et al., 2007, \apj, 659, L137
\bibitem[Carr \& Najita, 2008]{Carr08} Carr J.S., Najita J.R., 2008, Science, 319, 1504
\bibitem[Ceccarelli et al., 2005]{Cecca05} Ceccarelli C., Dominik C., Caux E., et al., 2005, \apj, 631, L81
\bibitem[Cernicharo et al., 1990]{Cerni90} Cernicharo J., Thum C., Hein H., et al., 1990, \aap, 231, L15
\bibitem[Cernicharo et al., 1994]{Cerni94} Cernicharo J., Gonz\'alez-Alfonso E., et al., 1994, \apj, 432, L59
\bibitem[Cernicharo et al., 1996]{Cerni96} Cernicharo J., Bachiller R., Gonz\'alez-Alfonso E., 1996, \aap, 305, L5
\bibitem[Cernicharo et al., 2006]{Cerni06} Cernicharo, J., Goicoechea J., Pardo J.R., Asensio-Ramons A., 2006, \apj, 642, 940
\bibitem[Chapillon et al., 2008]{Chapillon08} Chapillon E., Guilloteau S., Dutrey A., Pi\'etu V., 2008, \aap, 488, 565
\bibitem[Dominik et al., 2005]{Dominik05} Dominik C., Ceccarelli C., Hollenbach D., Kaufman M., 2005, \apj, 635, L85
\bibitem[Doucet et al., 2007]{Doucet07}Doucet C., Habart E., Pantin E., et al., 2007, \aap, 470, 625
\bibitem[Dubernet et al., 2009]{Duber09}Dubernet M.L., Daniel F., et al., 2009, \aap, 497, 911 
\bibitem[Faure et al., 2007]{Faure07} Faure A., Crimier N., Ceccarelli C., et al., 2007, \aap, 472, 1029
\bibitem[Gonz\'alez-Alfonso \& Cernicharo, 1993]{Gonza93} Gonz\'alez-Alfonso E., Cernicharo J., 1993, \aap, 279, 506
\bibitem[Gonz\'alez-Alfonso \& Cernicharo, 1997]{Gonza97} Gonz\'alez-Alfonso E., Cernicharo J., 1997, \aap, 322, 938
\bibitem[Green et al., 1993]{Green93} Green S., Maluendes S., McLean A.D., 1993, \apjs, 85, 181
\bibitem[Lagage et al., 2006]{Lagage06}Lagage P.O., Doucet C., Pantin E., et al., 2006, Science, 314, 621
\bibitem[Mathis \& Whiffen, 1989]{Mathis89} Mathis J.S., Whiffen G., 1989, \apj, 342, 808
\bibitem[Pavlyuchenkov et al., 2007]{Pavl07} Pavlyuchenkov Ya, Henning Th., Wiebe D., 2007, \apj, 669, L101
\bibitem[Pinte et al., 2006]{Pinte06} Pinte C., M\'enard F., Duchene G., Bastien P., 2006, \aap, 459, 797
\bibitem[Pinte et al., 2007]{Pinte07} Pinte C., Fouchet L., M\'enard F., et al. 2007, \aap, 469, 963
\bibitem[Pinte et al., 2008]{Pinte08} Pinte C., Padgett D.L., M\'enard F., et al. 2008, \aap, 489, 633
\bibitem[Pinte et al., 2009]{Pinte09} Pinte C., Harries T.J., Min, et al., 2009, \aap, 498, 967
\bibitem[Salyk et al., 2008]{Salyk08} Salyk C., Pontoppidan K., Blake G.A., et al., 2008, \apj, 676, L49
\bibitem[Willacy, 2007]{Willacy07} Willacy K., 2007, \apj, 660, 441
\bibitem[Woitke et al., 2009]{Woitke09} Woitke P., Thi W.F., Kamp I., et al., 2009, \aap 501, L5
\end{thebibliography}
\end{document}